\documentclass[12pt]{article}
\usepackage[a4paper, margin=1in]{geometry}
\usepackage{graphicx} 
\usepackage{amsmath}
\usepackage{booktabs}
\usepackage{multirow}
\usepackage{hyperref}
\usepackage{bm}
\usepackage{array}
\usepackage[symbol]{footmisc}
\usepackage{setspace}

\usepackage{tikz}
\usetikzlibrary{arrows,decorations.pathmorphing,backgrounds,positioning,fit,petri,shapes.geometric}
\tikzset{
    >=stealth',
    punkt/.style={
           rectangle,
           rounded corners,
           draw=black, very thick,
           text width=6.5em,
           minimum height=2em,
           text centered},
    pil/.style={
           ->,
           thick,
           shorten <=2pt,
           shorten >=2pt,}
}

\usepackage[
    backend=biber,
    natbib,
    style=authoryear,
    maxcitenames = 2,
    maxbibnames = 99,
    uniquelist=false,
    isbn=false,
    dashed=false,
    date=year,
  ]{biblatex}
\DefineBibliographyExtras{english}{}
\DeclareFieldFormat{urldate}{(accessed on #1)}
\patchcmd{\bibsetup}{\interlinepenalty=5000}{\interlinepenalty=10000}{}{}

\title{Autoregressive hidden Markov models for high-resolution animal movement data}

\author{}
\date{}

\begin{document}

\maketitle

\begin{center}\vspace{-3em}
    {\large Ferdinand V.\ Stoye$^{1*}$, Annika Hoyer$^1$, Roland Langrock$^2$}
\end{center}

\begin{center}
    $^1$ Biostatistics and Medical Biometry, Medical School OWL, Bielefeld University, Bielefeld, Germany\\
    $^2$ Statistics and Data Analysis, Department of Business Administration and Economics, Bielefeld University, Bielefeld, Germany
\end{center}

\vspace{1em}

\footnotetext[1]{\texttt{ferdinand.stoye@uni-bielefeld.de}}

\begin{abstract}
    \noindent New types of high-resolution animal movement data allow for increasingly comprehensive biological inference, but method development to meet the statistical challenges associated with such data is lagging behind. 
    In this contribution, we extend the commonly applied hidden Markov models for step lengths and turning angles to address the specific requirements posed by high-resolution movement data, in particular the very strong within-state correlation induced by the momentum in the movement. 
    The models feature autoregressive components of general order in both the step length and the turning angle variable, with the possibility to automate the selection of the autoregressive degree using a lasso approach.
    In a simulation study, we identify potential for improved inference when using the new model instead of the commonly applied basic hidden Markov model in cases where there is strong within-state autocorrelation. 
    The practical use of the model is illustrated using high-resolution movement tracks of terns foraging near an anthropogenic structure causing turbulent water flow features.
\end{abstract}


\doublespacing

\section{Introduction}\label{sec1}

\begin{figure}[!htb]
\begin{center}
\includegraphics*[width=0.9\textwidth]{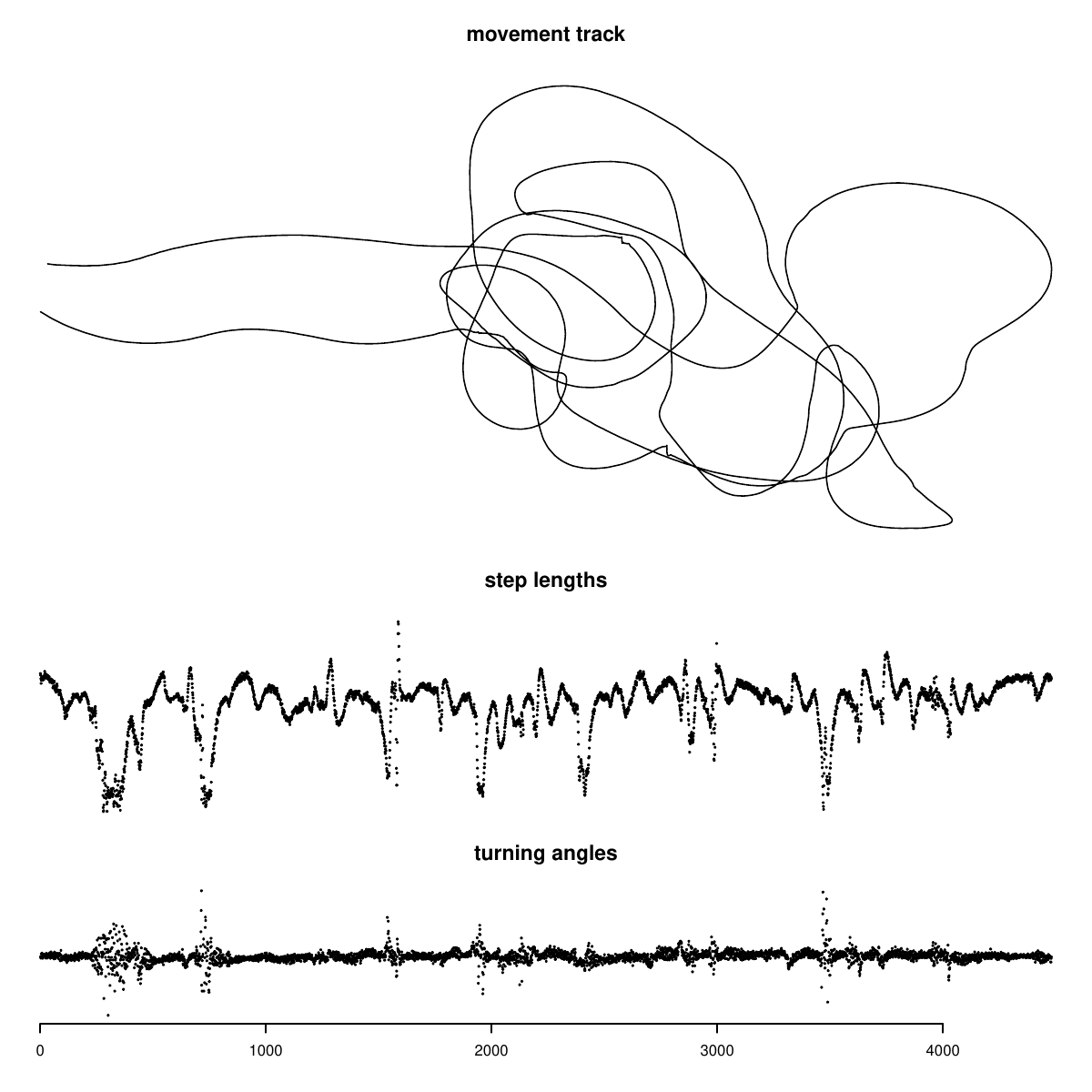}
\caption{Movement track and bivariate time series of step lengths and turning angles of a 
tracked tern (seabird) using aerial drone observations at 30 Hz (units are omitted).}
\end{center}
\label{fig:tern}
\end{figure}

High-resolution movement data, e.g.\ with a sampling resolution of 1~Hz or even higher, holds vast potential for ecological inference: behavioural modes and highly agile manoeuvres such as foraging attempts can be more accurately identified, social interactions and predator-prey encounters can be revealed, and the effects of environmental stimuli can be reliably estimated \parencite{nathan2022big}. Figure \ref{fig:tern} displays one such time series, showing step lengths and turning angles derived from the track of a tern species, a surface-foraging seabird, observed at 30~Hz \parencite{schwalbe_data}. 
Troughs in the time series of step lengths indicate short-lived foraging manoeuvres such as hovering or shallow plunge dives. 
The data-driven identification of such behaviours from movement data, and associated inference relating the behavioural process to internal and external covariates, is often conducted using \textit{hidden Markov models} (HMMs). 
The most common HMM formulation assumes each observed pair of step length and turning angle to be generated by one of $N$ possible underlying behavioural modes, and that these behavioural modes evolve according to an $N$--state Markov chain \parencite{langrock2012flexible,mcclintock2020uncovering,glennie2023hidden}.

However, a basic HMM would assume the observations to be conditionally independent, given the state sequence --- or, in other words, that within a state the observations are independent and identically distributed. For time series with coarse resolutions, say with locations measured on an hourly basis, the consequences of this simplifying dependence assumption will typically be negligible, in particular as a minor lack of fit with respect to capturing within-state serial correlation will hardly alter the decoding of behavioural modes. 
In contrast, for the high-resolution data shown in Figure \ref{fig:tern}, it is clear that corresponding models will exhibit a rather substantial lack of fit: the bird does seem to switch between two (or more) somewhat persistent behavioural modes, which induce correlation in the time series, but also within a behavioural mode both the sequence of steps and also the sequence of turning angles show high serial correlation, as these metrics vary gradually due to the linear and the angular momentum in the movement. 
With the increased availability of such high-resolution movement data, it is now crucial to more carefully evaluate possible negative consequences of the conditional independence assumption and, where necessary, to use more flexible model formulations capturing the within-state serial correlation.

In this contribution, we develop a flexible class of HMMs for steps and turns that incorporates within-state autoregressions and that is specifically designed to meet the requirements of high-resolution animal movement data. 
Our work extends a similar endeavour by \textcite{Lawler2019} in four ways: i) not only the step lengths but also the turning angles are modelled using within-state autoregression; ii) the step length distribution is modelled assuming a constant coefficient of variation; iii) all autoregressive components are modelled using general lag $p$; iv) lasso regularisation is used to select the order of the autoregression automatically. 
We further use simulations to gauge the consequences of using models with simpler dependence structures in scenarios with high within-state serial correlation.

The remainder of this article is organised as follows. In Section \ref{sec2}, we introduce the autoregressive HMM tailored to the analysis of high-resolution animal movement data. 
Subsequently, in Section \ref{sec3}, we investigate the estimation performance, in particular concerning consistency and (numerical) stability, in a simulation study. A case study using 1 Hz movement data of sea-foraging terns is provided in Section \ref{sec4}. We conclude the article with a discussion of our findings in Section \ref{sec5}.

\section{Model formulation}\label{sec2}

In movement ecology, the sequence of movement metrics considered --- for ease of interpretation and modelling often taken to be the step lengths and turning angles between successive locations --- is commonly modelled conditional on an underlying, non-observable sequence of states. The states are then typically interpreted as proxies of behavioural modes such as resting, foraging, or travelling. The corresponding class of HMMs involves i) an $N$--state Markov chain $\{ S_t\}_{t=1,\ldots,T}$, here for simplicity assumed to be homogeneous, defined by its initial state distribution $\boldsymbol\delta = \bigl( \Pr(S_1 =1) ,\ldots, \Pr(S_1 =N)\bigr)$ and transition probability matrix $\boldsymbol\Gamma=(\gamma_{ij})$, $\gamma_{ij}=\Pr (S_{t}=j | S_{t-1} =i)$, and ii) suitable state-dependent distributions for the observed movement metrics \parencite{Zucchini_2016}.

In the following, the notation $f$ will be used as a general symbol denoting either a probability mass function or a density, depending on context and data. In the most basic HMM formulation, there are two main assumptions, the \textit{Markov property} and the \textit{conditional independence assumption}. 
The Markov property describes the dependence structure of the state process. In the basic setting, the distribution of states at time $t$ is completely determined by the state active at time $t-1$:
$$ f(s_t \mid s_1,\ldots,s_{t-1})= f(s_t \mid s_{t-1}).$$
The conditional independence assumption fixes the state-dependent process $\{ X_t\}_{t=1,\ldots,T}$, whose realisations are the observed movement metrics, to be completely determined by the current state:
$$f(x_t\mid s_1,\ldots,s_{T},x_1\ldots,x_{t-1},x_{t+1},\ldots,x_T)=f(x_t\mid s_t).$$
In other words, conditional on the state currently active, previous (and future) states and observations do not comprise any additional information affecting the observation $x_t$.
When this assumption is not met, e.g.\ due to physical momentum in movement as for the tern data shown in Figure \ref{fig:tern}, the natural refinement is to allow additional dependence on past observations, i.e.\ autoregressive structures in the state-dependent process: 
$$f(x_t\mid s_1,\ldots,s_{T},x_1\ldots,x_{t-1},x_{t+1},\ldots,x_T)=f(x_t\mid s_t,x_{t-1},\ldots,x_{t-p}).$$
For additional flexibility, the order $p$ can also depend on the current state $s_t$. 
Corresponding model formulations were first popularised as \textit{Markov-switching autoregression} (or sometimes \textit{Regime-switching autoregression}) in econometric applications \parencite{hamilton1989new, goodwin1993business, boldin1996check, kim1999has}, but can be adapted to step lengths and turning angles as commonly analysed in movement ecology.

To formulate a flexible model for animal movement analyses, we consider the bivariate time series $\{ \mathbf{X}_t\}_{t=1,\ldots,T}$, $\mathbf{X}_t=(X_{t}^{\text{step}},X_{t}^{\text{turn}})$, with $X_{t}^{\text{step}}$ the step length and $X_{t}^{\text{turn}}$ the turning angle at time $t$. 
The non-negative, real-valued step lengths are defined as the (Euclidean) distances between consecutive location measurements, while the turning angles, defined in radians and hence with support $(-\pi,\pi]$, indicate the changes in movement direction as obtained from three consecutive locations. 

Conditional on the state, we assume the step length to follow a {gamma} distribution,
\begin{equation}\label{eq:step}
X_t^{\text{step}}\mid S_t=j \sim \Gamma\left(\mu_{t,j}^\text{step}, \sigma_{t,j}^\text{step}\right),
\end{equation}
where the state-dependent mean $\mu_{t,j}^\text{step}$ fluctuates around a steady-state mean $\mu_j^\text{step}$ according to an autoregressive process of order $p_j^{\text{step}}$,
\begin{equation*}
    \mu_{t,j}^{\text{step}} = \sum_{k=1}^{p_j^{\text{step}}}\phi_{j,k}^{\text{step}} x_{t-k}^{\text{step}} + \Bigl(1-\sum_{k=1}^{p_j^{\text{step}}}\phi_{j,k}^{\text{step}}\Bigr) \mu_j^\text{step}.
\end{equation*}
The state-dependent standard deviation $\sigma_{t,j}^\text{step}$ is calculated as $ \sigma_{t,j}^\text{step} = \omega_j \mu_{t,j}^{\text{step}} $, 
with the constant coefficient of variation $\omega_j$ a parameter to be estimated. 
To allow for step lengths equal to zero, an additional point mass at zero can be included. Other distributional families could be considered, but the gamma distribution seems preferable over natural alternatives such as the log-normal or the Weibull distribution --- the former typically does not fit step length data that well, and the latter cannot easily be parameterised in terms of a mean and standard deviation parameter.

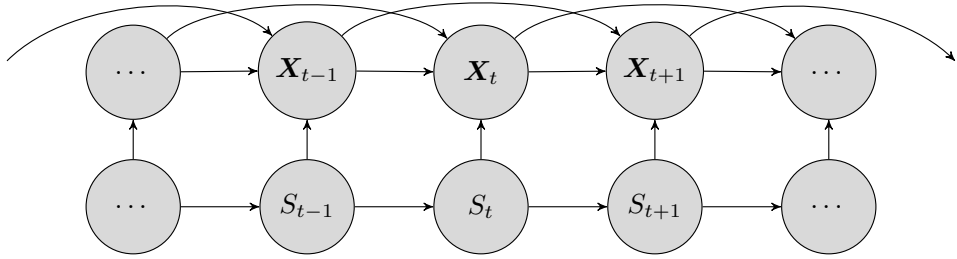
\begin{figure}[h]
    \centering
    \resizebox{0.9\linewidth}{!}{
    \begin{tikzpicture}
    \node(before)[circle, draw, minimum width=1.2cm,fill=gray!30!white]{\footnotesize\dots};
    \node(beforex)[circle, draw, minimum width=1.2cm,fill=gray!30!white, above=0.5cm of before]{\footnotesize\dots};
    \draw[->](before)--(beforex);
    
    \node(st-1)[circle, draw, minimum width=1.2cm, right=1cm of before,fill=gray!30!white]{\footnotesize$S_{t-1}$};
    \node(xt-1)[circle, draw, minimum width=1.2cm, above=0.5cm of st-1,fill=gray!30!white]{\footnotesize$\boldsymbol{X}_{t-1}$};
    \draw[->](st-1)--(xt-1);
    \draw[->](before)--(st-1);
    \draw[->](beforex)--(xt-1);
    
    \node(pastx)[left=1cm of beforex]{};
    \draw[->] (pastx) .. controls +(1,1) and +(-1,1) .. (xt-1);
    
    \node(st)[circle, draw, minimum width=1.2cm, right=1cm of st-1,fill=gray!30!white]{\footnotesize$S_{t}$};
    \node(xt)[circle, draw, minimum width=1.2cm, above=0.5cm of st,fill=gray!30!white]{\footnotesize$\boldsymbol{X}_{t}$};
    \draw[->](st)--(xt);
    \draw[->](st-1)--(st);
    \draw[->](xt-1)--(xt);
    \draw[->] (beforex) .. controls +(1,1) and +(-1,1) .. (xt);
    
    \node(st1)[circle, draw, minimum width=1.2cm, right=1cm of st,fill=gray!30!white]{\footnotesize$S_{t+1}$};
    \node(xt1)[circle, draw, minimum width=1.2cm, above=0.5cm of st1,fill=gray!30!white]{\footnotesize$\boldsymbol{X}_{t+1}$};
    \draw[->](st1)--(xt1);
    \draw[->](st)--(st1);
    \draw[->](xt)--(xt1);
    \draw[->] (xt-1) .. controls +(1,1) and +(-1,1) .. (xt1);
    
    \node(after)[circle, draw, minimum width=1.2cm, right=1cm of st1,fill=gray!30!white]{\footnotesize\dots};
    \node(afterx)[circle, draw, minimum width=1.2cm, above=0.5cm of after,fill=gray!30!white]{\footnotesize\dots};
    \draw[->](st1)--(after);
    \draw[->](xt1)--(afterx);
    \draw[->](after)--(afterx);
    \draw[->] (xt) .. controls +(1,1) and +(-1,1) .. (afterx);
    
    \node(futurex)[right=1cm of afterx]{};
    \draw[->] (xt1) .. controls +(1,1) and +(-1,1) .. (futurex);
    \end{tikzpicture}
    }
        \caption{Visualisation of the model's dependence structure with AR(2) structure in the state-dependent process. Every observation $\boldsymbol{X}_t$ contains the bivariate information on step length and turning angle, $\boldsymbol{X}_t=(X_t^{\text{step}},X_t^{\text{turn}})$, the additional assumption of conditional independence of steps and turns, given the states, is not made explicit here.}
    \label{fig:ar(2)-hmm}
\end{figure}

For the turning angles, the circular nature of the variable needs to be taken into account. This can, for example, be achieved by assuming a {von Mises} state-dependent distribution,
\begin{equation}
X_t^{\text{turn}}\mid S_t=j \sim \text{von Mises}\left(\mu_{t,j}^{\text{turn}}, \kappa_j^{\text{turn}}\right),
\label{eq:turn}
\end{equation}
where we formulate, for each state $j=1,\ldots,N$, an autoregressive process of order $p_j^{\text{turn}}$ on the mappings of the turning angles to their corresponding values on the unit circle \parencite{langrock2012flexible}:
\begin{equation*}
\mu_{t,j}^{\text{turn}}=\text{Arg}\Biggl(\sum\limits_{k=1}^{p_j^{\text{turn}}}\phi_{j,k}^{\text{turn}} \exp\left(i\, x_{t-k}^{\text{turn}}\right) + \Bigl( 1-\sum\limits_{k=1}^{p_j^{\text{turn}}}\phi_{j,k}^{\text{turn}}\Bigr)  \exp\left(i\, \mu_j^{\text{turn}}\right)\Biggr)
\label{eq:state-dep-turning-angle}.
\end{equation*}
Conditional on the state, we assume the gamma-distributed step length and the von Mises-distributed turning angle to be independent of each other.
For both the step lengths and the turning angles, we assume $\phi_{j,k}^{\cdot}\in [0,1]$, $\sum_{k=1}^{p_j}\phi_{j,k}^{\cdot}\le 1$, for the autoregressive parameters within state $j$, $j=1,\ldots,N$. We illustrate the overall dependence structure for the case $p_j=2$ (for both variables) in Figure \ref{fig:ar(2)-hmm}.
\textcite{ailliot2006some} showed consistency and asymptotic normality of the maximum likelihood estimator of Markov-switching autoregressive models with gamma-distributed observations, while \textcite{ailliot2015non} showed the consistency in case of von Mises-distributed observations. 
These results serve as a theoretical foundation of our model.

The model described above features $N\times(N-1)+4\times N+\sum_{j=1}^N (p_j^{\text{step}}+p_j^{\text{turn}})$ parameters, with the three summands corresponding to the state-switching probabilities, the state-dependent means, coefficients of variations and precisions, and the autoregressive coefficients, respectively. The initial state distribution $\boldsymbol\delta$ can be taken as the stationary distribution implied by $\boldsymbol\Gamma$.
Parameter estimation can be conducted by maximising the conditional likelihood, for each state $j$ conditioning on the first $p_j$ observations. To automate the choice of the state-dependent autoregressive order $p_j$, we additionally include a lasso penalty on the autoregressive coefficients, resulting in the \textit{partially penalised conditional likelihood}
\begin{align}\label{eq:like} 
\nonumber \mathcal{L} & = \boldsymbol\delta \mathbf{P}(x_1^{\text{step}},x_1^{\text{turn}}) \boldsymbol{\Gamma} \mathbf{P}(x_2^{\text{step}},x_2^{\text{turn}}) \cdot \ldots \cdot \boldsymbol{\Gamma} \mathbf{P}(x_T^{\text{step}},x_T^{\text{turn}}) \mathbf{1}^{t} \\ 
& \qquad - \lambda 
\biggl( 
\sum_{j=1}^N \sum_{k=1}^{p_j^{\text{step}}} \vert \phi_{j,k}^{\text{step}} \vert + 
\sum_{j=1}^N \sum_{k=1}^{p_j^{\text{turn}}} \vert \phi_{j,k}^{\text{turn}} \vert
\biggr), 
\end{align}
with a complexity penalty $\lambda \geq 0$.  The diagonal matrices $\mathbf{P} (x_t^{\text{step}},x_t^{\text{turn}})$ comprise the products of the state-dependent gamma and von Mises densities as implied by (\ref{eq:step}) and (\ref{eq:turn}) on the diagonal, for each state $j$ replacing the initial $p_j$ state-dependent densities by ones. 
To obtain a differentiable objective function, we follow \textcite{oelker2017uniform} and approximate the $L_1$ norm $| \cdot |$ in the penalty by $\sqrt{ (\cdot + \epsilon)^2}$, with a small positive $\epsilon$, in the implementation. 
To still be able to obtain estimates equal to zero and thus to allow for automated degree selection, we round the estimated autoregressive parameters to three decimal places (cf.\ \citealp{hambuckers2018understanding,otting2022regularized}).

\section{Simulation study}\label{sec3}

To evaluate the proposed model framework in a controlled setting, we conducted a simulation study. 
We begin by introducing the parameter choices of the data-generating models, before continuing with considerations related to estimator consistency, to the accuracy of the state decoding, and to numerical stability. 
Finally, we investigate the behaviour of the estimated autoregressive parameters regarding correct degree selection during penalisation as proposed in (\ref{eq:like}).

\subsection{Setup}\label{sec:3.1}

We simulate data from basic and autoregressive $2$-state HMMs with autoregression degrees 1--3 in the state-dependent distributions of both the step lengths and the turning angles. 
The models as proposed in (\ref{eq:step}) and (\ref{eq:turn}), again with varying autoregression degrees, are then re-fitted to the simulated data.
In total, we execute $16$ different simulation configurations --- four different models from which we draw data, four different models that are then fitted to the simulated data. 
We run $250$ iterations for each configuration, in
every iteration sampling $T=2000$ data points and performing global decoding using the \textit{Viterbi algorithm} \parencite{viterbi1967error} after model fitting.

For the true state-dependent distributions and autoregressive parameters, we choose values that result in distributions with medium overlap to emulate a realistic scenario. The parameter choices can be found in Table 1 in the Supplementary Material to this article.
All analyses are conducted in \texttt{R version 4.4.1} \parencite{R-core}, utilizing the package \texttt{Rcpp} \parencite{rcpp, r-rcpp} to be able to implement parts of the log-likelihood in \texttt{C++} and thus speed up computation. 
To enable full reproducibility of the simulation results, all simulation code is accessible at \url{https://gitlab.ub.uni-bielefeld.de/stoyef/autoregressivehmm}.

\subsection{Estimation without lasso penalisation}

\begin{figure}[htbp]
    \centering
    \includegraphics[width=0.9\linewidth]{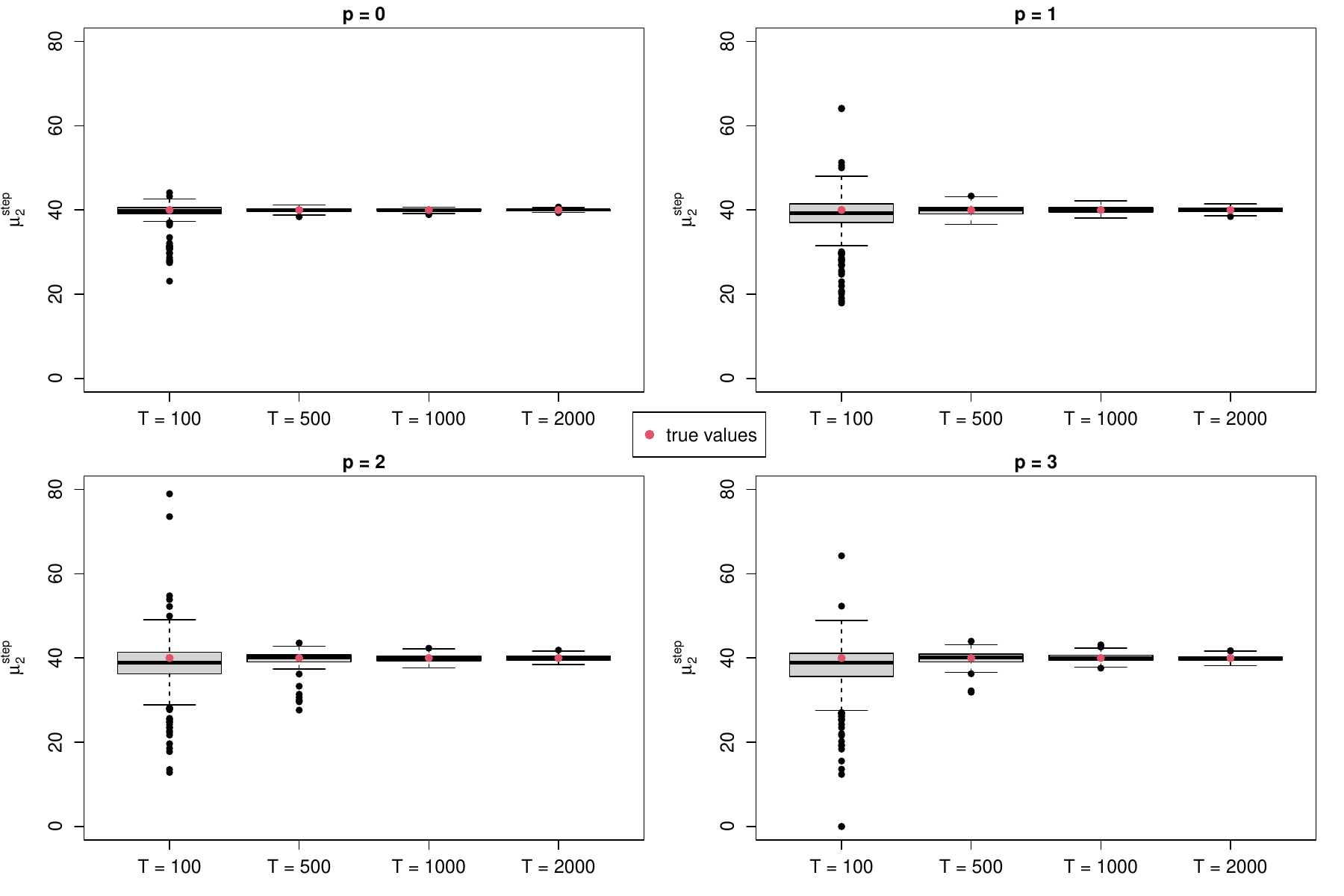}
        \caption{Boxplots of the estimates of the parameter $\mu_2^\text{step}=40$ when varying the autoregressive degree $p$ (one panel for each $p=0,1,2,3$) and the number of data points $T$.}
    \label{fig:consistency}
\end{figure}

We initially fit our models to the simulated data fixing $\lambda=0$ in the likelihood calculation, thus not incorporating any penalisation. 
To empirically validate consistency, we choose $T\in\{100,500,1000,2000\}$ and investigate the estimators' performance when fitting the correctly specified models. 
Figure \ref{fig:consistency} illustrates the consistency for one example parameter being estimated. Results for the other parameters are comparable.
The standard errors of the estimators increase with increasing order $p$ of the autoregression, which is to be expected given the additional model complexity. 

\begin{figure}[!htb]
    \centering
    \includegraphics[width=\linewidth]{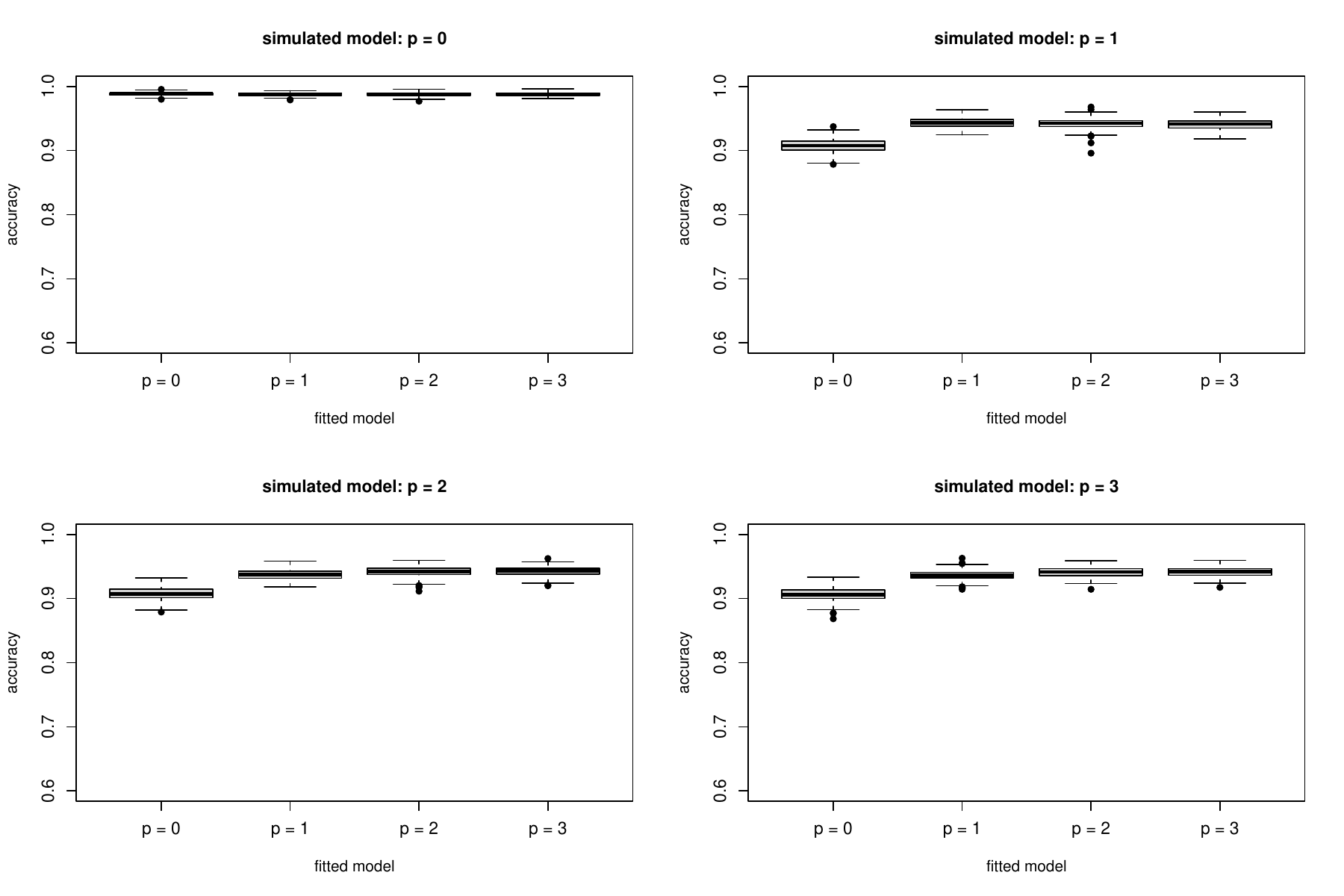}
    \caption{Boxplots of global decoding accuracies using the Viterbi algorithm, obtained using different models in the simulation (represented by the four different panels) and then fitting both the correctly and also incorrectly specified models to the simulated data ($p=0,1,2,3$ as indicated on the horizontal axes).}
    \label{fig:performance}
\end{figure}

The central motivation for developing the autoregressive model structure is that the standard HMM architecture is not able to successfully capture within-state autocorrelation as commonly found in high-resolution movement data. Such a lack of fit can have negative consequences in particular on the accuracy of the state decoding.
Thus, to assess the potential benefit of the proposed models in this regard, we fit models with $p\in\{0,1,2,3\}$ to data generated as described in Section \ref{sec:3.1} and use the Viterbi algorithm \parencite{viterbi1967error} to globally decode the states. 
Figure \ref{fig:performance} shows boxplots of the resulting classification accuracies when compared to the known true states. 

We find that incorporating within-state autoregression in the model can substantially improve the state decoding performance when the true data do exhibit within-state autocorrelation (from around 91\% to around 95\% accuracy).
Notably, the concrete choice of $p$ does not have that large an effect on the decoding accuracies: for example, when the data are simulated using $p=3$, then fitting models with $p=1$ already substantially improves the state decoding compared to the model without autoregression.
Finally, we find that using an autoregressive structure, i.e.\ fitting models with $p>0$, does not notably decrease the state decoding accuracy when the data do in fact not exhibit within-state correlation.

It is important to note that the absolute differences in the state decoding accuracies naturally depend on the configuration of the state-dependent distributions, as the decoding will in any case be very accurate when the states are clearly separated.
To analyse if the relative performance differences also depend on this overlap in the underlying densities, we vary the amount of overlap in the state-dependent distributions. 
Figures 1 and 2 in the Supplementary Material show two such alternative scenarios and illustrate that the relative differences in the accuracies remain very similar when changing the underlying distributional parameters.


Model stability quantifies the proportion of simulation runs in which a global 
optimum has been reached. 
In our default simulation setting, using $T=2000$, the global optimum was found in all runs. 
When decreasing the number of data points in each data set, this stability diminishes. 
Table \ref{tab:stability} shows the resulting proportions of global optima for $T=100$. 
While all models still perform satisfactory, a small decline in stability can be observed, when increasing the autoregressive degree in the data-generating process. 

\begin{table}[!htpb]
    \centering
    \begin{tabular}{cccccc}\toprule
    &&\multicolumn{4}{c}{\textbf{fitted model}}\\
    && $p=0$ & $p=1$ & $p=2$ & $p=3$\\\midrule
    \multirow{4}{*}{\textbf{simulated model}} 
         &$p=0$& $0.936$ & $0.864$ & $0.836$ & $0.908$ \\
         &$p=1$& $0.844$ & $0.848$ & $0.856$ & $0.816$   \\
         &$p=2$& $0.812$ & $0.844$ & $0.787$ & $0.820$ \\
         &$p=3$& $0.760$ & $0.784$ & $0.848$ & $0.776$ \\\bottomrule
    \end{tabular}
    \caption{Proportions simulation runs in which the global optimum of the log-likelihood for the given data was identified ($T=100$).}
    \label{tab:stability}
\end{table}

In summary, including within-state autoregressive structures in the model is feasible and can notably improve the decoding accuracy (and, as a consequence, also the forecasting performance, which we did not investigate here as it is usually not the focus of analyses of animal movement data). Autoregression orders higher than $p=1$ will typically not be necessary to obtain this improvement even if the true data-generating process does have a higher-order dependence structure. The consequences of an unnecessary inclusion of autoregressive structure are for the most part negligible, though numerical stability may slightly decrease.

\subsection{Choosing the autoregression degree with lasso}\label{sec:3.5}

We now briefly investigate the estimation performance when using partial lasso penalisation as in (\ref{eq:like}). 
In this context, we are primarily interested in whether this approach can automatically determine the optimal autoregressive degree $p$, and if it impacts the decoding accuracy of the models.

For each of $250$ simulation runs, we simulate $T=2000$ observations from the default simulation setting with true autoregression degree $p=2$, and then fit models considering possible penalties $\lambda$ from a grid. In each run, the optimal $\lambda$ is selected using the Akaike Information Criterion (AIC), the Bayesian Information Criterion (BIC), and the global decoding accuracy, allowing for a maximum autoregressive degree of five. 
We specify the autoregressive degree of a fitted model by rounding the estimated autoregressive parameters to three decimal places and then determining $p$ as the maximum time lag for which the associated estimated coefficient is non-zero (cf.\ \citealp{hambuckers2018understanding}, \citealp{otting2022regularized}).

\begin{figure}[!htb]
    \centering
    \includegraphics[width=\linewidth]{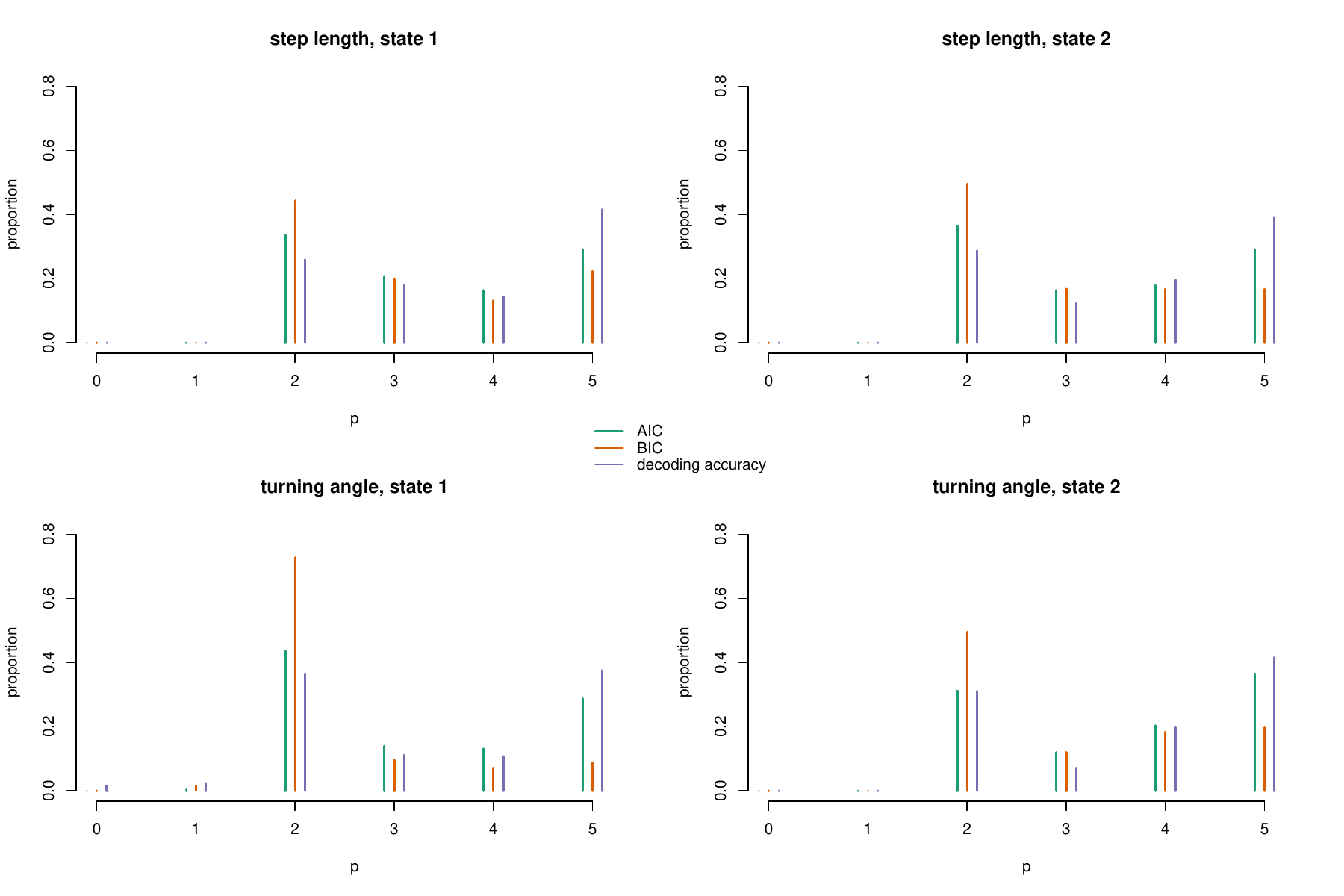}
    \vspace{-4em}
        \caption{Overview of proportion of runs in which autoregression degrees 0--5 were chosen (with $p=2$ the correct value) using lasso penalisation, for each of the state-dependent distributions and using three different methods for selecting the penalty $\lambda$ (AIC, BIC, and global decoding accuracy).
    }
    \label{fig:compare_true_to_chosen_p}
\end{figure}

Figure \ref{fig:compare_true_to_chosen_p} displays the results, in particular showing which proportion of runs resulted in the correct degree $p=2$, for each of the four state-dependent distributions involving autoregressive terms. In almost all runs, the degree $p$ was either correctly chosen, with the BIC yielding the best results in that respect, or was estimated to be larger than necessary. 
A comparison of the estimated optimal penalties $\lambda$ or the estimated effective degrees of freedom supports this statement (cf.\ Figures 4 and 5 in the Supplementary Material). 
We further find that the accuracy of the global decoding accuracy is not strongly affected by the choice of $\lambda$, with only very large complexity penalties --- that were in fact not chosen in our simulations --- resulting in a notable decline of the accuracy (cf.\ Figure 6 in the Supplementary Material).
Overall, these results suggest that penalisation can support the otherwise rather cumbersome choice of the autoregression degrees. The degree may however be overestimated.

\section{Application to tern data}\label{sec4}

To demonstrate and test the autoregressive HMM in a case study, we apply it to high-resolution tern movement data obtained from videos recorded by a drone hovering over a floodtide wake in Northern Ireland (see \citealp{lieber2021bird}, for more details on this data set). The terns adapt their foraging behaviour to exploit the water turbulences caused by the anthropogenic structure. We restrict our example analysis to those tracks from the original data set that comprised $> 1000$ observations collected at the original 30~Hz resolution, downsample these tracks to 1~Hz and pre-process them using the \texttt{R} package \texttt{moveHMM} \parencite{moveHMM} to obtain step length and turning angle variables. One of the tracks was already shown in Figure \ref{fig:tern} and all are freely available from \citet{schwalbe_data_source}. As all tracks originate from a single drone deployment and hence were subject to the same environmental conditions, it can reasonably be expected that foraging behaviour across tracks is homogeneous, such that we resort to a complete pooling approach when analysing the different tracks using a joint model \parencite{Zucchini_2016}. This does of course imply that we neglect any potential individual heterogeneity, which we argue is adequate here given that this aspect is not the focus of our example analysis.

\begin{figure}[!htb]
    \centering
    \includegraphics[width=0.9\linewidth]{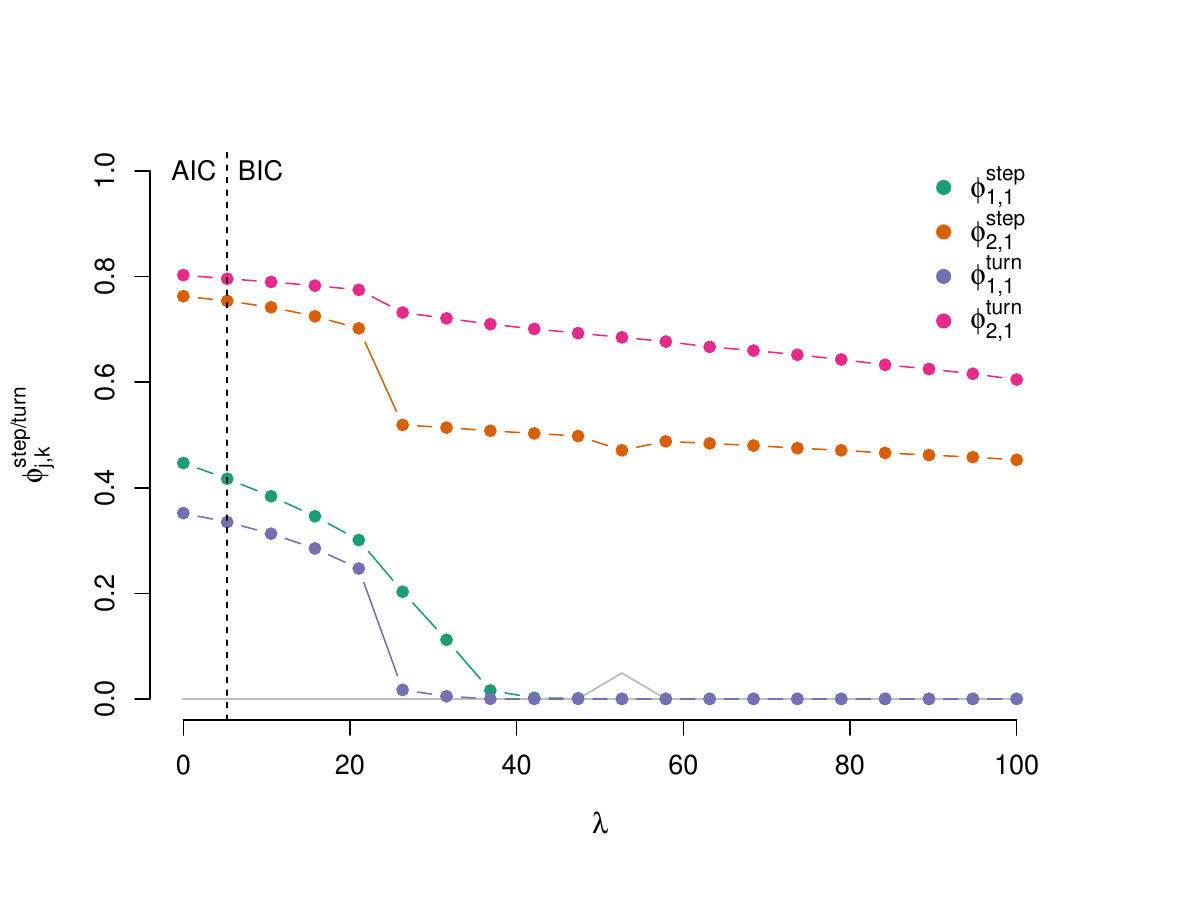}
    \vspace{-2em}
    \caption{Trajectories of the estimates of the autoregression coefficients in the 2-state HMM fitted to the terns' steps and turns, for increasing complexity penalty $\lambda >0$. For clarity of presentation, the legend comprises only those coefficients that differ notably from zero for some $\lambda$ and should be included in the model based on information criteria.
    }
    \label{fig:tern_auto_params}
\end{figure}

The model described above, with $N=2$ states, was fitted using maximisation of the partially penalised conditional likelihood (\ref{eq:like}), allowing for a maximum autoregressive degree $p=5$ in each state and variable. Figure \ref{fig:tern_auto_params} displays the trajectories of the autoregression coefficients $\phi_{j,k}$ for increasing complexity penalty $\lambda$, indicating successful lasso-type variable selection. 
Both AIC 
and BIC select a model featuring autoregressive terms of degree one in each state and variable. This confirms the observation from our simulations that potential higher degrees of autocorrelation in the data can be captured fairly effectively by first-degree autoregressive components.

Based on the autoregression degrees selected when using the information criteria, we re-fit the model with $\lambda=0$, i.e.\ no penalisation, to retain full interpretability of the estimated parameters \citep{zhao2021defense}. The resulting parameter estimates are as follows:
\begin{gather*}
    \bm{\hat\delta} = (0.401, 0.599); \quad \bm{\hat\Gamma} = \begin{pmatrix}
        0.792&0.208\\ 0.139&0.861\end{pmatrix};\\
    \bm{\hat\mu}^{\text{step}}=(647.882, 898.314); \quad \bm{\hat\sigma}^{\text{step}}=(225.487, 62.210);\\
    \bm{\hat\mu}^{\text{turn}}=(0.044,0.119); \quad \bm{\hat\kappa}^{\text{turn}}=(3.049, 18.376);\\
    \hat\phi_{1,1}^{\text{step}}=0.465;\quad
    \hat\phi_{2,1}^{\text{step}}=0.791;\quad 
    \hat\phi_{1,1}^{\text{turn}}=0.348;\quad
    \hat\phi_{2,1}^{\text{turn}}=0.799.
\end{gather*}

In both states, the mean turning angle is estimated to be close to zero, implying movement according to a correlated random walk. State 1 is associated with moderately long steps that show high variability and also comparably little persistence in the movement direction. State 2 in contrast involves exclusively large steps in combination with a very high directional persistence. The latter is thus indicative of transit flights, whereas the former is mostly associated with active search behaviour \citep{lieber2021bird}. 
For both movement speed (as measured by $\hat\phi_{j,1}^{\text{step}}$, $j=1,2$) and curvature of the flight path (as measured by $\hat\phi_{j,1}^{\text{turn}}$, $j=1,2$), the estimates indicate moderate serial correlation within the first state and relatively high serial correlation in the second state.

When manually exploring the different autoregressive degrees (rather than using the lasso), considering all possible models with autoregressive degree $\boldsymbol{p}\in\{0,1,2,3\}^4$, we find that the BIC favours a model with $p=1$ in state 1 and $p=3$ in state 2 for both variables, i.e.\ a higher degree in state 2 compared to the outcome when using the lasso approach (see Table 4 in the Supplementary Material).  
The observed discrepancy between the automated degree selection and manual exploration of the model space can be explained by the partial penalisation with lasso. When applying the lasso, we obtain a different likelihood than by fixing the autoregressive degrees in advance and forcing $\lambda=0$. 
This may result in deviations during model selection. The estimated parameters of the best model found by manual exploration are reported in Section B.1 in the Supplementary Material.

Model selection criteria provide relative measures for the goodness-of-fit of candidate models, but cannot reveal any potential absolute lack of fit of the selected models. 
Such additional model checking is commonly performed using pseudo-residuals, which assess the individual observations relative to the fitted model and the complete time series; for the hypothetical ``true'' model, these residuals are standard normally distributed, and vice versa any deviation from normality indicates a lack of fit \parencite{Zucchini_2016}. 
Figure \ref{fig:tern_pseudores} compares kernel density estimates of the pseudo-residuals obtained under the models selected above --- a basic HMM, the autoregressive HMM chosen using lasso, and the autoregressive HMM chosen via manual exploration ---  to the density of a standard normal distribution. 
We find that the models that include autoregressive components both seem to provide a satisfactory fit, in contrast to the basic HMM without autoregressive components. This is confirmed by visually comparing tracks simulated from the different models considered (cf.\ Figure 7 in the Supplementary Material): in particular, the basic HMM fails to capture the persistence in the curvature of the movement.

In summary, including within-state autoregressive components in the standard HMM for steps and turns data leads to a notable improvement in model fit in the case study considered, in particular with a view on the model being able to produce realistic synthetic tracks. 
While (faster) automated degree selection using lasso leads to satisfying results, the manual degree selection here still outperformed the lasso approach when consulting model selection criteria. 
In general, the selection of the optimal autoregressive degree poses a challenging problem that should not be reduced to a naive comparison of AIC or BIC \parencite{pohle2017selecting}. 
Computing pseudo-residuals and simulating data from the fitted models can support decision-making in this regard.

\section{Discussion}\label{sec5}

In this contribution, we explored the potential incorporation of flexible autoregressive models in the state-dependent process of hidden Markov models for step lengths and turning angles as nowadays routinely analysed in movement ecology. There is an increasing need for such model formulations due to an increasing availability of data sets collected at high temporal resolutions, for which the commonly made assumption of conditional independence of observations within HMM states will typically be severely violated. To address this caveat, we tailored Markov-switching autoregression models of general degree as commonly used in econometric applications to the specific nature of steps and turns data, and propose an optional automatic degree selection using a lasso penalty on the autoregressive coefficients. 
 
In simulation settings where the true data-generating process does involve within-state autocorrelation, we found our model to perform notably better than basic HMM formulations, with the proportion of incorrectly decoded states decreasing roughly by a factor of two (in the settings we considered). A difference of this magnitude can clearly affect associated inference on the response of the behavioural state process to covariate effects, which is often the key focus of HMM analyses in movement ecology (see, e.g., \citealp{towner2016sex,grecian2018understanding,whittington2022towns}).
Notably, the state decoding performance of our approach is competitive even in scenarios where the true observation process does not involve any autocorrelation within states. In other words, we found that the approach can lead to substantial improvements in the state decoding accuracy and associated ecological inference, while the risk of a deterioration due to overfitting is limited. We also found that it will often not be necessary to consider autoregressive degrees higher than one --- even if there may be conceptual reasons for considering higher degrees based on physical considerations related to the momentum of movement --- as a simple first-order autoregression will typically already result in a notable improvement and capture much of the residual autocorrelation. 

In our case study on high-resolution tracking data collected for sea-foraging terns, we found a notable improvement in the goodness-of-fit when using the new model formulation  compared to the standard HMM formulation. We also illustrated that model selection, which is difficult already for more basic HMMs  \citep{pohle2017selecting}, is not made easier by the additional model complexity resulting from the consideration of autoregressive components. In particular, when using the convenient lasso-type approach to select the degree of the autoregressive components, we had to resort to a rather naive approach for estimating the effective degrees of freedom. With autoregressive components in HMM-type models, there is also an increased risk of numerical instability as these models effectively comprise two different options to capture serial correlation --- via the dynamics of the state process but also via the correlation within states --- leading to potentially challenging identifiability (or rather estimability) issues. Our overall conclusion is thus that the new model formulation and possible variations thereof should be used cautiously in practice, nevertheless the high serial correlation typically found in high-resolution data certainly warrants the consideration of HMMs with autoregressive components, as the improvements in model fit and consequently also the validity of the ecological inference can be substantial.

\paragraph{Acknowledgements:}
The authors are grateful to Dr.\ Lilian Lieber for her helpful comments on an earlier version of this paper.

\paragraph{Data availability statement:} 
All programming code used in the simulation study and
the model application, as well as the full data set used in
the application, is available at \url{https://gitlab.ub.uni-bielefeld.de/stoyef/autoregressivehmm}. The data set was originally published in \textcite{schwalbe_data_source}.

\printbibliography


\end{document}


\maketitle

\begin{center}\vspace{-2em}
    $^1$ Biostatistics and Medical Biometry, Medical School OWL, Bielefeld University, Bielefeld, Germany\\
    $^2$ Statistics and Data Analysis, Department of Business Administration and Economics, Bielefeld University, Bielefeld, Germany
\end{center}

\appendix

\section{Additional simulation results}

\subsection{Parameter choices in simulation study}

\begin{table}[htbp]
    \centering
    \resizebox{\linewidth}{!}{
    \begin{tabular}{
    >{\centering\arraybackslash} m{2.75cm} 
    >{\centering\arraybackslash} m{1cm} 
    >{\centering\arraybackslash} m{12cm} 
    }\toprule
    variable & $p$ & values\\\midrule
        step length & $1$ & $\boldsymbol{\mu}^{\text{step}}=(20,40); \quad\boldsymbol{\sigma}^{\text{step}}=(5,7);\quad 
            \boldsymbol{\phi}^{\text{step}}=\begin{pmatrix}0.45\\0.55\end{pmatrix}
            $\\
        step length & $2$ & $\boldsymbol{\mu}^{\text{step}}=(20,40); \quad\boldsymbol{\sigma}^{\text{step}}=(5,7);\quad 
            \boldsymbol{\phi}^{\text{step}}=\begin{pmatrix}0.3&0.15\\0.4&0.15\end{pmatrix}
            $\\
        step length & $3$ & $\boldsymbol{\mu}^{\text{step}}=(20,40); \quad\boldsymbol{\sigma}^{\text{step}}=(5,7);\quad 
            \boldsymbol{\phi}^{\text{step}}=\begin{pmatrix}0.25&0.1&0.1\\0.35&0.1&0.1\end{pmatrix}
            $\\
        turning angle & 1 & $\boldsymbol{\mu}^\text{turn}=(0,0);\quad \boldsymbol{\kappa}^\text{turn}=(2,12);\quad \boldsymbol{\phi}^\text{turn}=\begin{pmatrix}0.5\\0.6\end{pmatrix}$\\
        turning angle & 2 & $\boldsymbol{\mu}^\text{turn}=(0,0);\quad \boldsymbol{\kappa}^\text{turn}=(2,12);\quad \boldsymbol{\phi}^\text{turn}=\begin{pmatrix}0.3&0.2\\0.4&0.2\end{pmatrix}$\\
        
        turning angle & 3 & $\boldsymbol{\mu}^\text{turn}=(0,0);\quad \boldsymbol{\kappa}^\text{turn}=(2,12);\quad \boldsymbol{\phi}^\text{turn}=\begin{pmatrix}0.3&0.1&0.1\\0.4&0.1&0.1\end{pmatrix}$\\\bottomrule
    \end{tabular}
    }
    \caption{True parameters in the simulation study. The autoregressive degree $p$ is assumed to be the same in step length and turning angle for all states. $\phi_{jk}$ of the matrices of autoregressive parameters is to be interpreted as the parameter for state $j$ in time $t-k$, $\forall t=p+1,\dots,T$.}
    \label{tab:pars_sim}
\end{table}

\noindent Table \ref{tab:pars_sim} shows the true parameter choices in the simulation study. To fully specify the models, we also report the stationary distribution of the state process, $\boldsymbol{\delta}=\left(0.5,0.5\right)$, 
as well as the t.p.m.\, 
$$\boldsymbol{\Gamma}=\begin{pmatrix}0.9&0.1\\0.1&0.9\end{pmatrix}.$$
Throughout our simulations, we use the stationary distribution as the initial distribution. 
We then assume the state process to start in the stationary distribution during model estimation. 

\subsection{Simulation performance}

\begin{figure}[htbp]
    \centering
    \includegraphics[width=\linewidth]{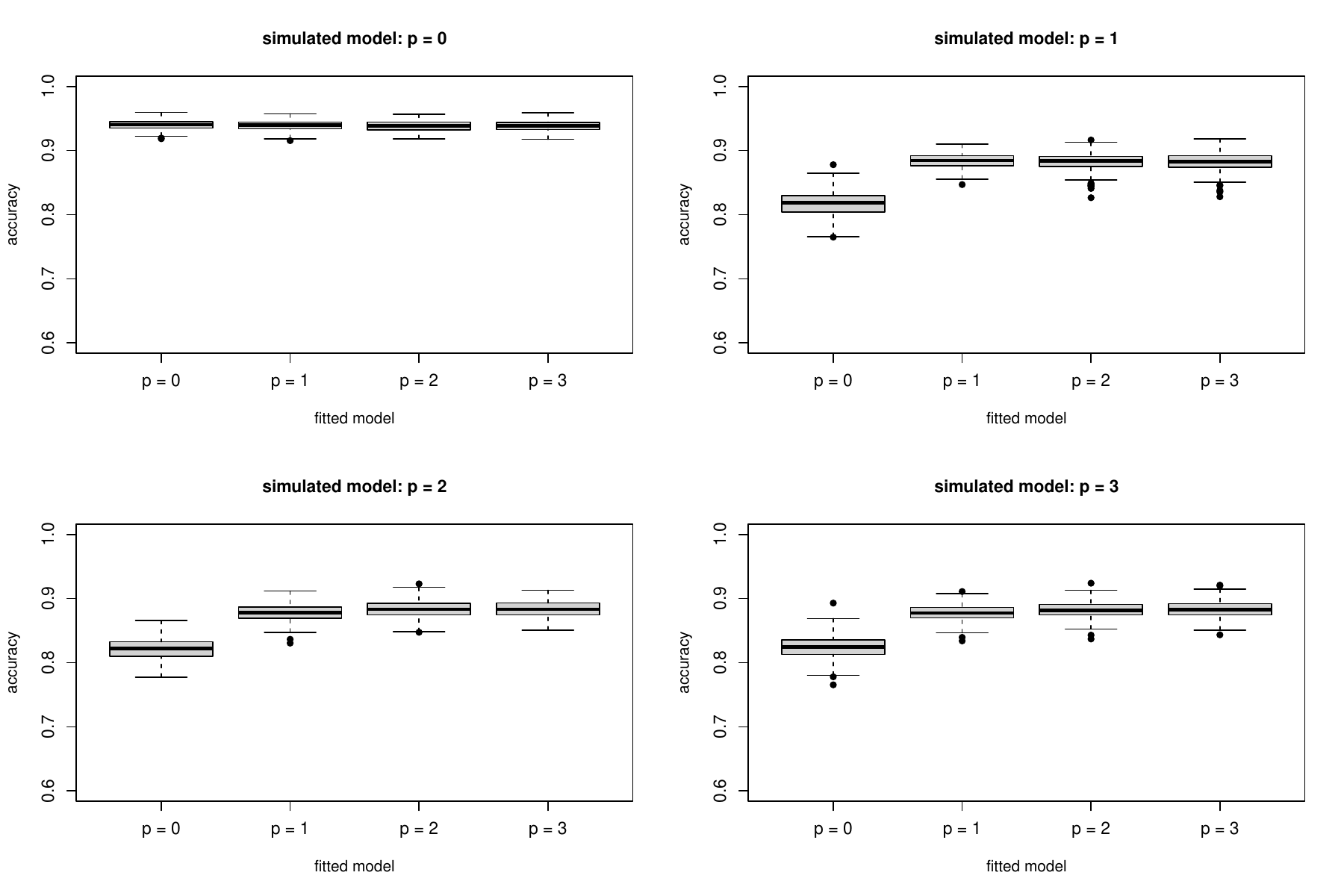}
    \caption{Boxplots of global decoding accuracies using the Viterbi algorithm for increased overlap in the state-dependent distributions ($\mu_1^\text{step}=25,\ \mu_2^\text{step}=35$).}
    \label{fig:performance_large_overlap_1}
\end{figure}

\begin{figure}[htbp]
    \centering
    \includegraphics[width=\linewidth]{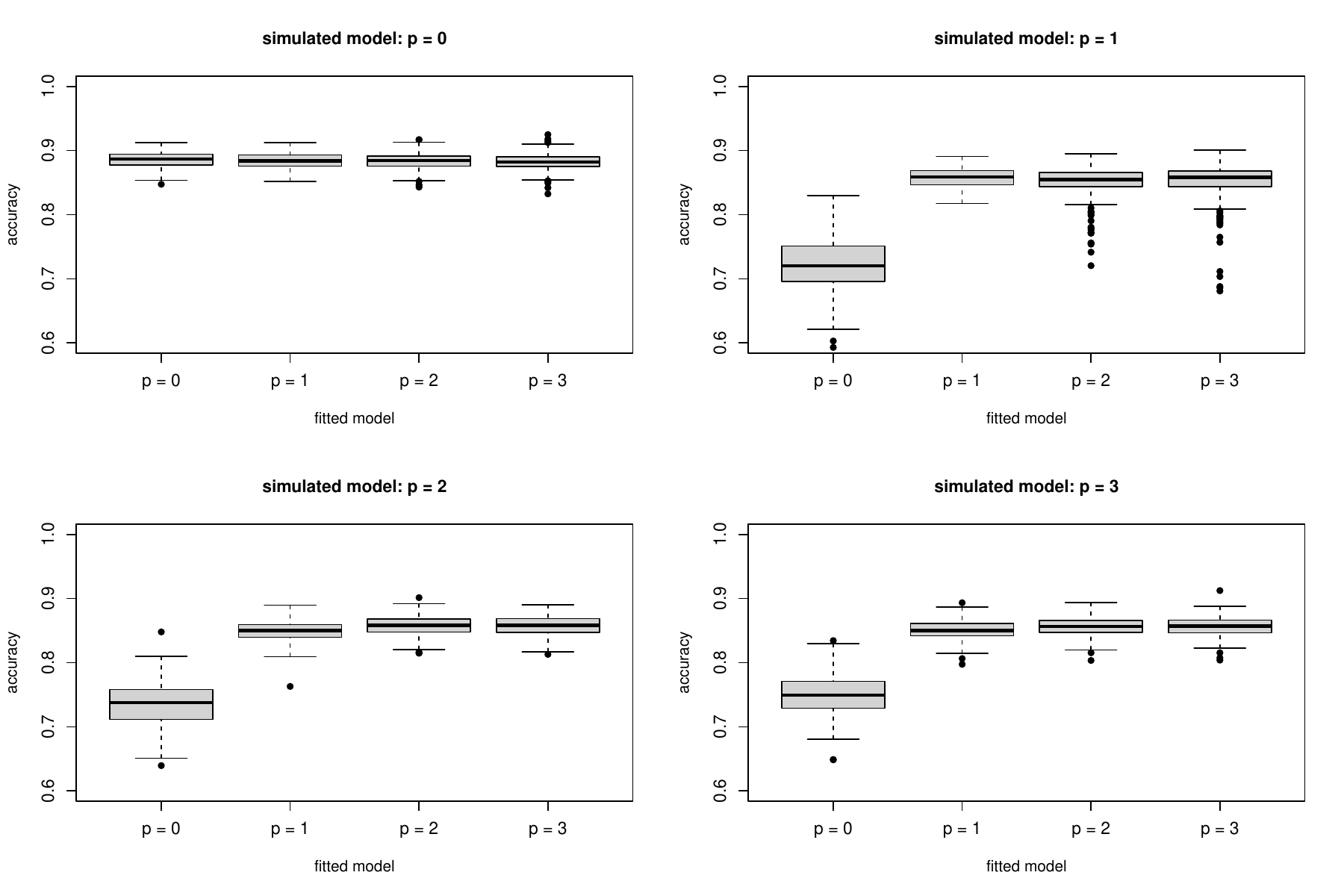}
    \caption{Boxplots of global decoding accuracies using the Viterbi algorithm for increased overlap in the state-dependent distributions ($\mu_1^\text{step}=25,\ \mu_2^\text{step}=30$).}
    \label{fig:performance_large_overlap_2}
\end{figure}

To compare the dependence of global decoding accuracies on the overlap between state-dependent distributions, we reduce the difference between $\mu_1^\text{step}$ and $\mu_2^\text{step}$ (ceteris paribus). Figure \ref{fig:performance_large_overlap_1} ($\mu_1^\text{step}=25,\ \mu_2^\text{step}=35$) and Figure \ref{fig:performance_large_overlap_2} ($\mu_1^\text{step}=25,\ \mu_2^\text{step}=30$) show that, while overall decoding accuracy is reduced for increased overlap, the difference in performance between basic and autoregressive HMMs persist.

\subsection{Stability}

Table \ref{tab:stability-250} and Table \ref{tab:stability-500} contain the proportions of global optima in the simulation with $T=250$, respective $T=500$. 

\begin{table}[H]
    \caption{Proportion of presumably global optima when carrying out 250 iterations for each model configuration when $n=250$.}
    \centering
    \begin{tabular}{cccccc}\toprule
    &&\multicolumn{4}{c}{\textbf{fitted model}}\\
    && $p=0$ & $p=1$ & $p=2$ & $p=3$\\\midrule
    \multirow{4}{*}{\textbf{simulated model}} 
         &$p=0$& $1.000$ & $0.960$ & $0.968$ & $0.972$ \\
         &$p=1$& $0.988$ & $0.964$ & $0.956$ & $0.964$ \\
         &$p=2$& $0.972$ & $0.968$ & $0.952$ & $0.944$ \\
         &$p=3$& $0.948$ & $0.976$ & $0.960$ & $0.952$ \\\bottomrule
    \end{tabular}
    \label{tab:stability-250}
\end{table}

\begin{table}[H]
    \caption{Proportion of presumably global optima when carrying out 250 iterations for each model configuration when $n=500$.}
    \centering
    \begin{tabular}{cccccc}\toprule
    &&\multicolumn{4}{c}{\textbf{fitted model}}\\
    && $p=0$ & $p=1$ & $p=2$ & $p=3$\\\midrule
    \multirow{4}{*}{\textbf{simulated model}} 
         &$p=0$& $1.000$ & $0.976$ & $0.988$ & $0.992$ \\
         &$p=1$& $1.000$ & $1.000$ & $0.984$ & $0.988$ \\
         &$p=2$& $0.996$ & $0.996$ & $0.972$ & $0.988$ \\
         &$p=3$& $0.996$ & $0.992$ & $0.992$ & $0.992$ \\\bottomrule
    \end{tabular}
    \label{tab:stability-500}
\end{table}

\subsection{Choice of complexity penalty}

\begin{figure}[H]
\caption{AIC and BIC of default simulation setting when varying $\lambda$, allowing a maximum autoregression degree of five.}
    \centering
    \includegraphics[width=\linewidth]{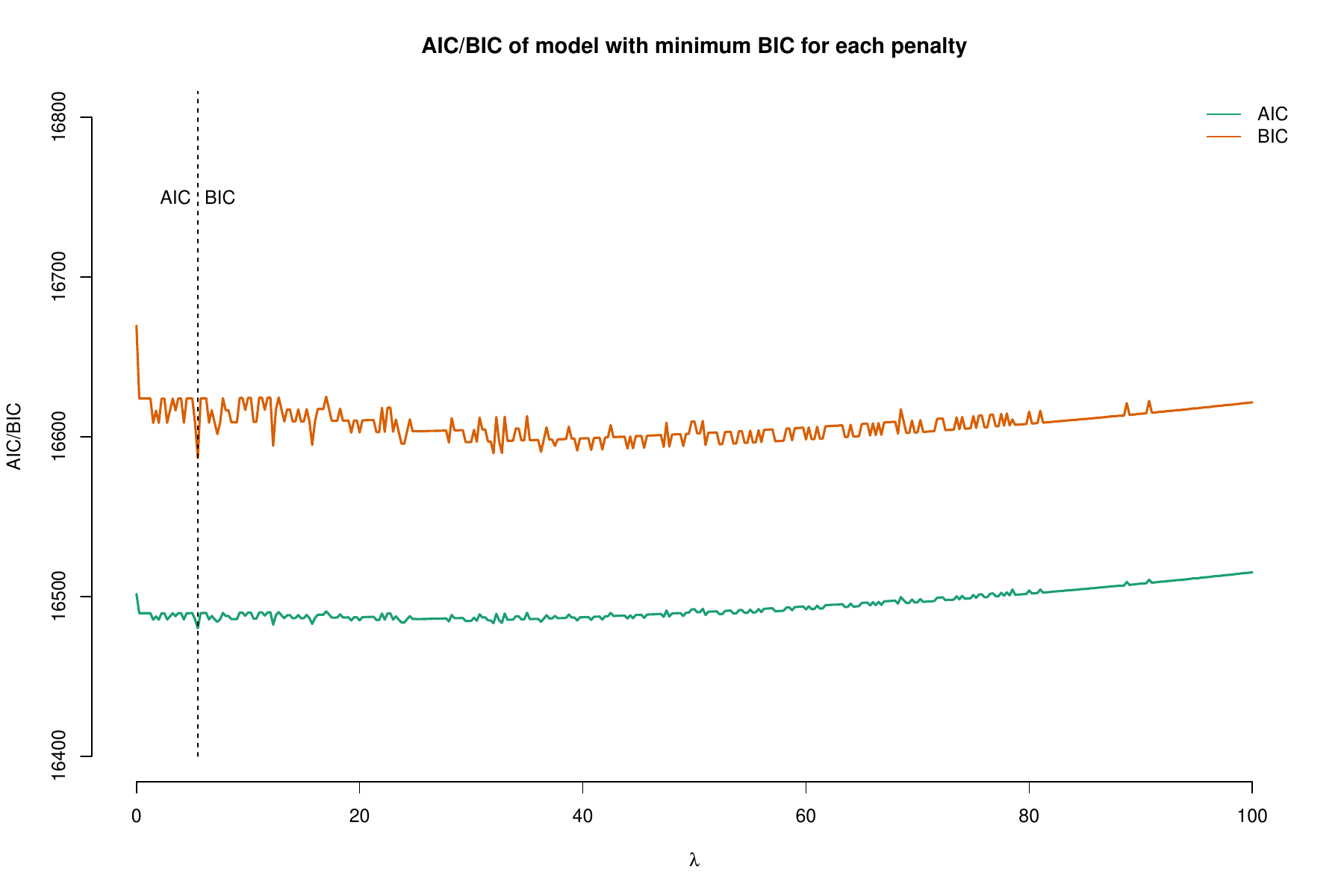}
    \label{fig:sim_aicbic_pen}
\end{figure}

Figure \ref{fig:sim_aicbic_pen} shows the development of AIC and BIC for increasing complexity penalty $\lambda\in [0,100]$ for a fixed simulated data set with true within-state autocorrelation of degree two. 
The fitted models allow for a maximum autoregressive degree of five in each variable and state. 
The vertical line indicates the optimal value of $\lambda$ regarding AIC and BIC. 

\begin{figure}[H]
    \caption{Histograms of chosen optimal complexity penalty $\lambda$ based on AIC, BIC and global decoding accuracy. 250 iterations on a grid of 24 values for $\lambda$ were computed in total.
    }
    \centering
    \includegraphics[width=\linewidth]{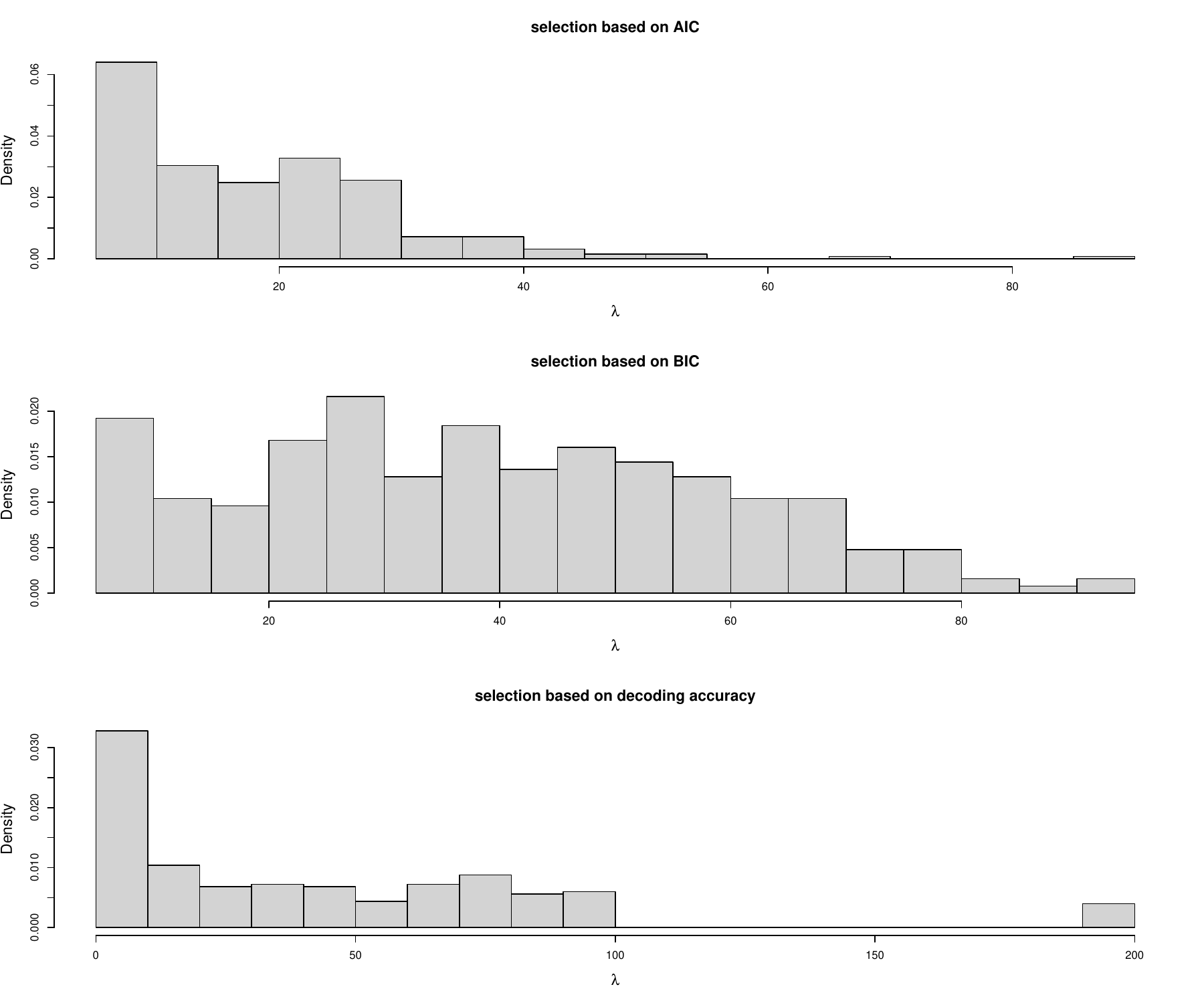}
    \label{fig:choice_of_lambda}
\end{figure}

Figure \ref{fig:choice_of_lambda} visualizes the distribution of optimal complexity penalties when selecting based on AIC, BIC, or global decoding accuracy. 
It is evident that the BIC chooses most conservatively. 

\begin{figure}[H]
    \caption{Histograms of effective degrees of freedom of models, selected based on AIC, BIC and global decoding accuracy. 250 iterations were computed in total.
    }
    \centering
    \includegraphics[width=\linewidth]{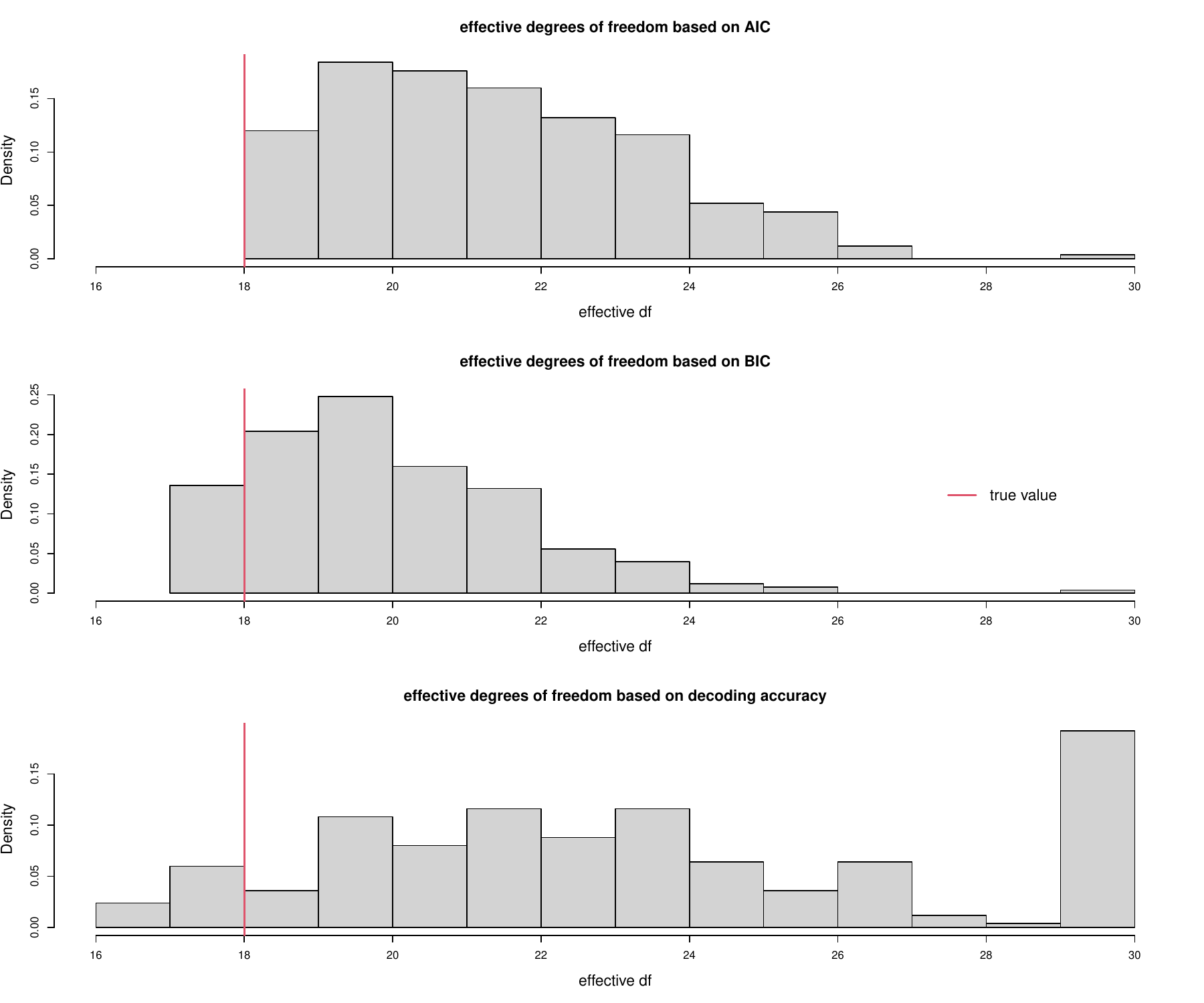}
    \label{fig:effective_df}
\end{figure}

Figure \ref{fig:effective_df} illustrates the systematic overestimation of effective degrees of freedom, compared to the true number of parameters in the underlying data-generating process. 
Again, the BIC reveals lower values than the other criteria, indicating it to be most suitable for model selection in our approach. 

\begin{figure}[H]
    \caption{Decoding accuracies of 100 lasso penalized models with varying complexity penalty $\lambda$.}
    \centering
    \includegraphics[width=0.9\linewidth]{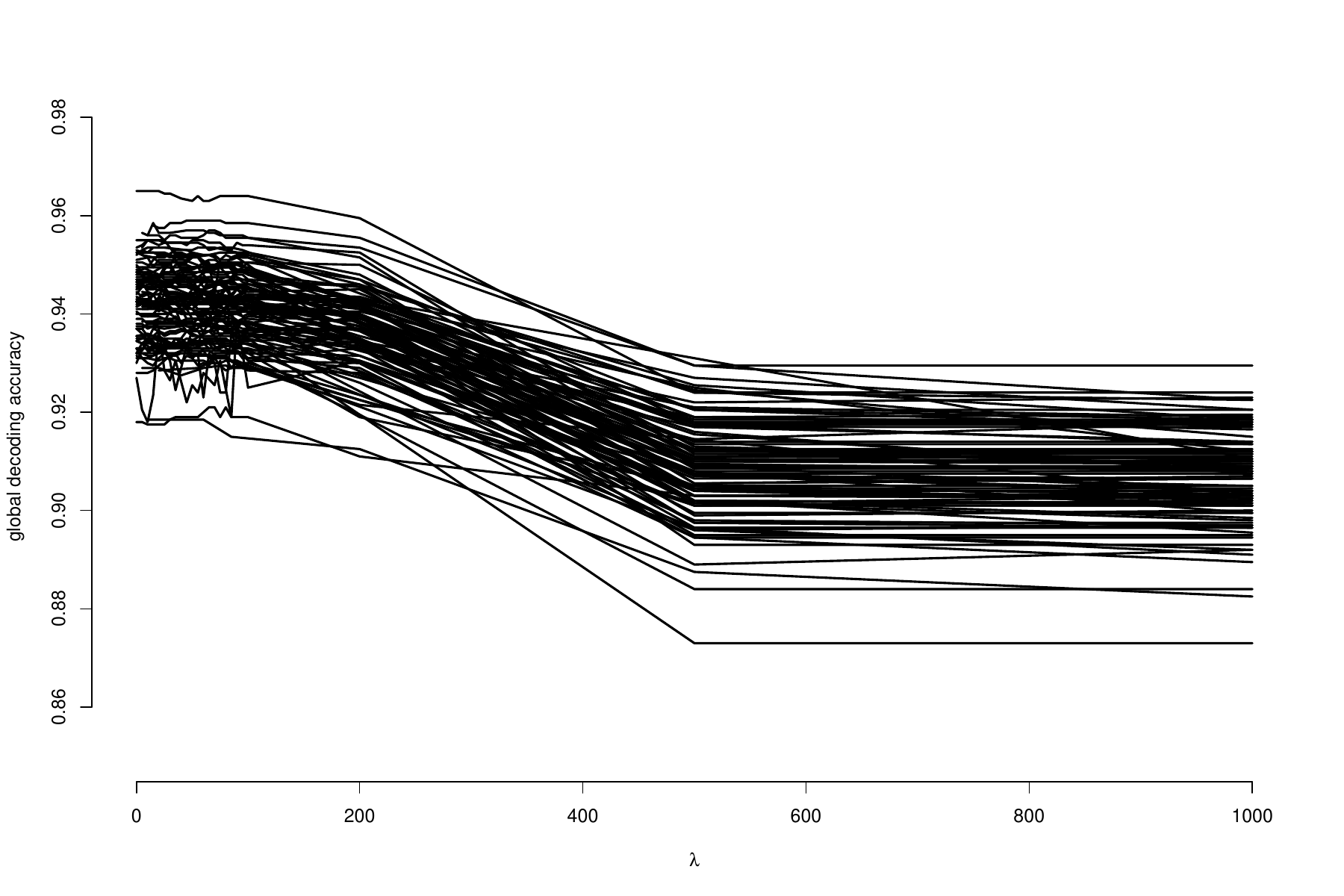}
    \label{fig:acc_different_lambda}
\end{figure}

Figure \ref{fig:acc_different_lambda} reveals the overall development of global decoding accuracies when increasing $\lambda$ for 100 different simulated data sets. 
While the accuracies remain fairly stable for low penalties, they decrease substantially if the penalties get large enough to delete autoregressive components from the model that are included in the true model formulation. 

\section{Additional application results}

\subsection{Additional parameter estimates of application to sea tern data}
\label{sec:app_params}

The estimated model parameters of the suggested model specification using manual exploration (within-state autoregression of first degree in state 1 and third degree in state 2) are:

\begin{align*}
    \boldsymbol{\hat\delta} &= (0.267, 0.733); \quad \boldsymbol{\hat\Gamma} = \begin{pmatrix}
        0.684&0.316\\ 0.115&0.885\end{pmatrix}\\
    \boldsymbol{\hat\mu}^{\text{step}}&=(563.308, 885.996); \quad \boldsymbol{\hat\sigma}^{\text{step}}=(212.256, 79.926)\\
    \boldsymbol{\hat\mu}^{\text{turn}}&=(0.048,0.121); \quad \boldsymbol{\hat\kappa}^{\text{turn}}=(2.509, 16.183)
\end{align*}
The estimated autoregressive parameters amount to 
\begin{gather*}
\hat\phi_{1}^{\text{step}}=0.487;\quad
\boldsymbol{\hat\phi}_{2}^{\text{step}}=(0.503,0.000,0.050);\\
\phi_{1}^{\text{turn}}=0.317;\quad
\boldsymbol{\hat\phi}_{2}^{\text{turn}}=(0.770,0.000,0.000).
\end{gather*}

\subsection{Model comparison}

Table \ref{tab:eval_tern_crit} compares different unpenalised model fits for the sea tern data using BIC. While the lasso penalisation selects the model with $\boldsymbol{p}=(1,1,1,1)$, the comparison of unpenalised models leads the the best result regarding BIC for $\boldsymbol{p}=(1,3,1,3)$.

\begin{table}[h]
    \centering
    \begin{tabular}{cccccc}\toprule
    &\multicolumn{5}{c}{\textbf{autoregressive degree}}\\
    $\boldsymbol{p}$ & (0,0,0,0) & (1,1,1,1) & (2,2,2,2) & {(1,3,1,3)} & (3,3,3,3) \\\midrule
     $\boldsymbol{\ell}$ & $-3668.65$ & $-3307.06$ &  $-3242.76$ & $-3181.89$ & $-3175.46$ \\
     \textbf{BIC} & $\phantom{-}7399.51$ & $\phantom{-}6701.21$ & $\phantom{-}6597.49$ & $\phantom{-}6475.76$ & $\phantom{-}6487.76$ \\\bottomrule
    \end{tabular}
    \caption{Comparison of selected model performances for the sea tern data (1~Hz) according to the BIC, calculated based on unpenalised model fitting. The degree $(1,1,1,1)$ was chosen using the lasso-based approach.}
    \label{tab:eval_tern_crit}
\end{table}

\subsection{Simulated movement tracks using estimated model parameters}

\begin{figure}[!htb]
    \centering
    \includegraphics[width=\linewidth]{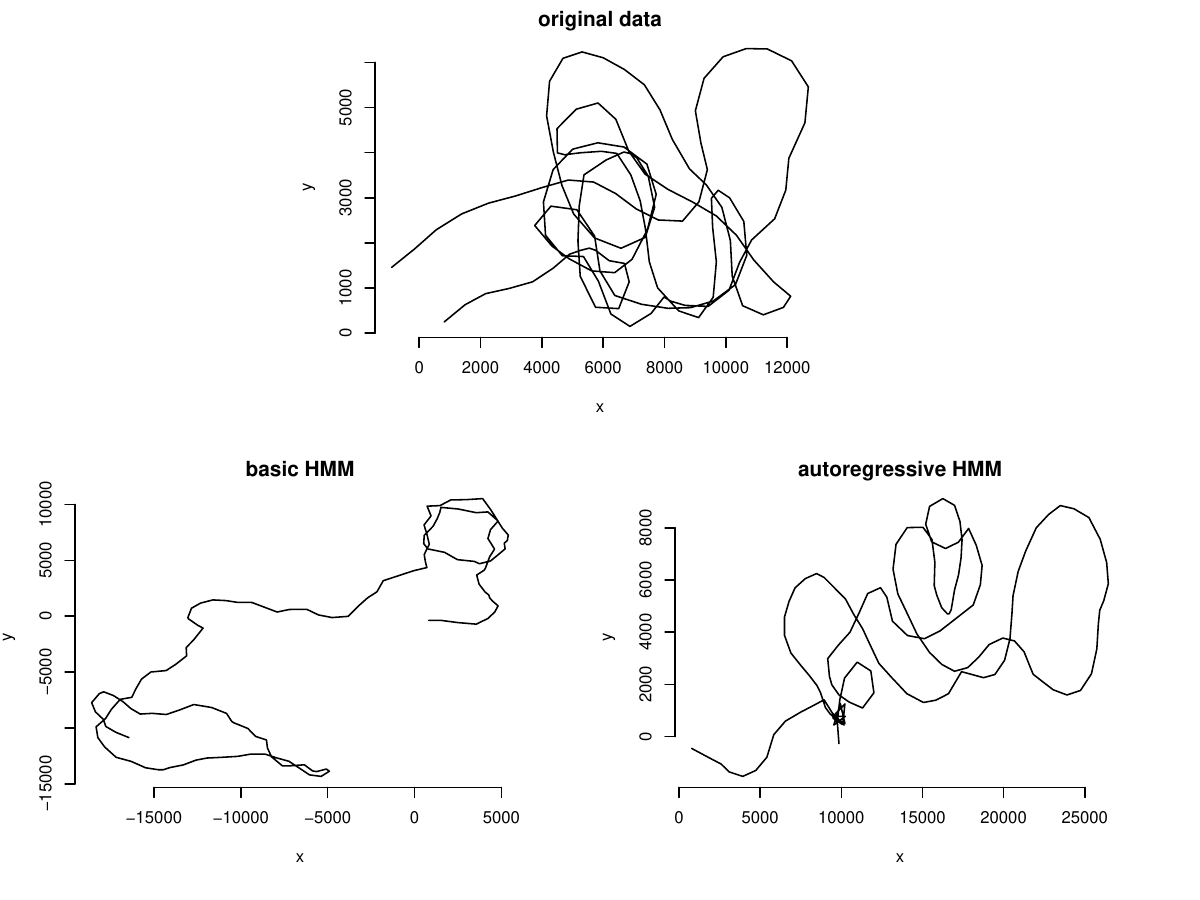}
    \caption{Comparison of real data (top) to data simulated from a fitted basic HMM (bottom left) and a fitted autoregressive HMM (bottom right). In the latter case, the choice of $\lambda\approx5.3$ corresponds to the best-scoring model based on information criteria.
    }
     \label{fig:tracks_comparison}
\end{figure}

To illustrate model adequacy, Figure \ref{fig:tracks_comparison} shows one of the original movement tracks (top) as well as simulated tracks from a basic HMM (bottom left) and the autoregressive HMM selected by the information criteria (bottom right) when applying automated lasso-based degree selection. 
The latter is able to produce more pronounced circles than the basic model formulation, leading to visually more realistic synthetic movement tracks. 

